\documentclass[twocolumn,prl,showpacs,floatfix]{revtex4}
\usepackage{color}
\usepackage{graphicx}%
\usepackage{psfrag}
\usepackage{dcolumn}
\usepackage{amsmath}
\usepackage{latexsym}

\begin{document}

\def\cr{\textcolor{red}}
\def\bea{\begin{eqnarray}}
\def\eea{\end{eqnarray}}
\def\beq{\begin{equation}}
\def\eeq{\end{equation}}
\def\f{\frac}
\def\k{\kappa}
\def\sx{\sigma_{xx}}
\def\sy{\sigma_{yy}}
\def\sxy{\sigma_{xy}}
\def\e{\epsilon}
\def\ve{\varepsilon}
\def\ex{\epsilon_{xx}}
\def\ey{\epsilon_{yy}}
\def\exy{\epsilon_{xy}}
\def\be{\beta}
\def\D{\Delta}
\def\h{\theta}
\def\r{\rho}
\def\a{\alpha}
\def\s{\sigma}
\def\kb{k_B}
\def\la{\langle}
\def\ra{\rangle}
\def\nn{\nonumber}
\def\bu{{\bf u}}
\def\bn{\bar{n}}
\def\br{{\bf r}}
\def\up{\uparrow}
\def\dn{\downarrow}
\def\S{\Sigma}
\def\dg{\dagger}
\def\d{\delta}
\def\p{\partial}
\def\l{\lambda}
\def\G{\Gamma}
\def\g{\gamma}
\def\kv{\bar{k}}
\def\ha{\hat{A}}
\def\hv{\hat{V}}
\def\hg{\hat{g}}
\def\hG{\hat{G}}
\def\hTT{\hat{T}}
\def\noi{\noindent}
\def\a{\alpha}
\def\d{\delta}
\def\p{\partial}
\def\r{\rho}
\def\xv{\vec{x}}
\def\rv{\vec{r}}
\def\fv{\vec{f}}
\def\ov{\vec{0}}
\def\vv{\vec{v}}
\def\la{\langle}
\def\ra{\rangle}
\def\e{\epsilon}
\def\o{\omega}
\def\g{\gamma}
\def\th{\hat{t}}
\def\uh{\hat{u}}
\def\break#1{\pagebreak \vspace*{#1}}
\def\f{\frac}
\def\uu{\vec{u}}
\def\vf{\vec{f}}

\def\vfm{\vec{f}^m}
\def\vfe{\vec{f}^e}
\def\fmx{{f}^{m}_{x}}
\def\fmy{{f}^{m}_{y}}
\def\fmox{{f}^{m}_{0x}}
\def\fmoy{{f}^{m}_{0y}}
\def\vxo{v^0_{x}}
\def\vyo{v^0_{y}}
\def\km{k_{m}}
\def\xm{x_m}
\def\ym{y_m}
\def\mux{{\mu}_{x}}
\def\muy{{\mu}_{y}}
\def\rom{\rho_m}
\def\pb{P_b}
\def\pu{P_u}
\def\vrm{\vec {r}_{m}}
\def\vjb{\vec {J}_{b}}
\def\vju{\vec {J}_{u}}
\def\won{\omega_{on}}
\def\woff{\omega_{off}}
\def\bwon{\bar{\omega}_{on}}
\def\bxm{{\bar{x}}_{m}}
\def\bym{{\bar{y}}_{m}}
\def\w{\omega}
\def\hi{\phi_i}
\def\hj{\phi_j}
\def\hip{\phi_{i+1}}
\def\hin{\phi_{i-1}}
\def\hf{\frac{1}{2}}
\def\n{\eta}
\def\t{\tau}
\def\vh{\vec{h}}
\title{Spontaneous helicity of a polymer with side-loops confined to a cylinder}
\author{Debasish Chaudhuri}
\email{chaudhuri@amolf.nl}
\author{Bela M. Mulder}
\email{mulder@amolf.nl}
\affiliation{
FOM Institute AMOLF,
Science Park 104,
1098XG Amsterdam,
The Netherlands
}

\date{\today}

\begin{abstract}
Inspired by recent experiments on the spatial organization of bacterial chromosomes,
we consider a type of ``bottle brush'' polymer consisting of a flexible backbone chain,
to which flexible side loops are connected. We show that such a model with an open linear
backbone spontaneously adopts a helical structure with
a well-defined pitch when confined to small cylindrical volume. This helicity persists over
a range of sizes and aspect-ratios of the cylinder, provided the packing fraction of the
chain is suitably large. We analyze this results in terms of the interplay between the
effective stiffness and actual intra-chain packing effects caused by the side-loops in
response to the confinement. For the case of a circular backbone, mimicking e.g.\ the
\emph{E. coli} chromosome, the polymer adopts a linearized configuration of two parallel helices
connected at the cylinder poles.
\end{abstract}
%\pacs{87.19.lp}{Pattern formation: activity and anatomic}
%\pacs{05.40.-a,05.40.Fb, 61.25.H-}

\maketitle

%\section{Introduction}
That confinement can dramatically influence the properties of non-ideal
polymers has already been appreciated theoretically for a long time. Scaling
arguments suggest that when a confining dimension $D$ becomes smaller than
the natural size of a linear polymer, typically given by its radius of gyration $R$,
the polymer, rather than behaving like a single coherent `blob' of size $R$, effectively
becomes a string of `blobs' of size $D$ \cite{Gennes1979}, signalling a cross-over to
lower-dimensional behavior. The experimental exploration of this regime, however, is difficult using
synthetic polymers, whose typical radii of gyration are in the nm regime and moreover tend to be highly
polydisperse in length~\cite{polymerhandbook}.
\begin{figure}[h]
\begin{center}
\includegraphics[height=7cm]{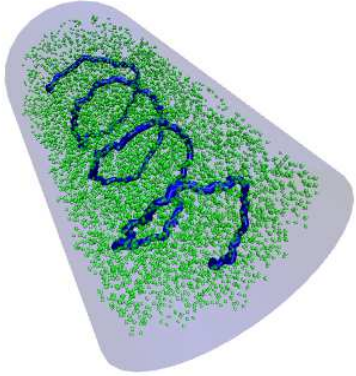} %{r_helix.eps}
\caption{(Color online)
Helical equilibrium structure for a backbone chain of length $l_b=200\s$ [black (blue) thick line] to
which side-loops of length $l_s=40 \s$ are attached at each backbone monomer. The polymer is confined
within a cylinder of length $L=50.75 \s$ and diameter $D=29.5 \s$. The side-loop monomers are shown as
transparent gray (green) beads.}
\label{helix-1}
\end{center}
\end{figure}
Fortunately nature provides us with an important class of model polymers in the form of DNA. To start with DNA
is produced with almost perfect length control. A well-studied example in vitro is purified
$\lambda$-phage DNA which contains exactly 48502 basepairs and has a length of ~16.5 $\mu$m
and a radius of gyration of ~2.1 $\mu$m. Using biochemical tools this length can be scaled to multiples of this unit. Moreover, DNA can readily be fluorescently labeled. These two properties of DNA were exploited in a number of recent experiments aimed at studying the behavior of polymers confined to nanofluidic
channels~\cite{Bonthuis2008, Tang2010, Uemura2010}.

More importantly, in vivo DNA in the form of
chromosomes is almost invariably strongly confined. The paradigmatic case is the 1.5
$\mathrm{mm}$ long circular chromosome of the bacterium \emph{E. coli}, which is confined to
a cylindrical volume of diameter $\sim 0.8 \mu\mathrm{m}$ and length varying between
$2 - 4 \mu\mathrm{m}$. Over the past few years there has been increasing interest in
understanding the details of the chromosomal configuration in \emph{E. coli} and its implications
for the, as yet far from fully understood, mechanism of sister-chromosome segregation prior
to division \cite{Bates2005,Jun2006,Wiggins2010}.
Strikingly, recent microscopic observations have shown that bacterial chromosomes
can exhibit a helical spatial density distribution within the cellular confinement,
with a pitch-length in the order of a fraction of the cell length~\cite{Berlatzky2008, Butan2011}.

Here our aim is to explore to what extent the physics of highly confined polymers by itself provides clues to such novel global configurations.
As a first step in that direction it is necessary to take into account that the structure of a typical bacterial chromosome is not
simply that of a closed linear polymer chain. The combined effects of supercoiling, due to a globally maintained
under-twist of the DNA, the action of a number of chromosome remodelling proteins, or even electrostatic ``zippering''
can cause loops in the DNA \cite{StevenB2006}, indicating that a more complex structural picture is needed.
Although in reality the dynamics of the formation and the statistics of such loops are likely to be highly complex,
almost certainly involving polydispersity in loop sizes and topological entanglements, we chose to focus on arguably
the simplest model of a polymer ``dressed'' by a cloud of loops, i.e.\ a backbone chain to which side-loops of equal size are attached at a regular spacing. The same model has recently also been discussed by Reiss et al. \cite{Reiss2011}, who have studied its
behavior in the absence of confinement effects. This type of polymer model is similar to the so-called ``bottle-brush'' polymers,
extensively studied by Binder and co-workers~\cite{Wang2010,Hsu2010a,Theodorakis2011}. The latter work has shown that
such polymers develop a local resistance to bending due to the entropic repulsion between the side chains.
This  effective stiffening, combined with intra-chain packing effects within the cylindrical confinement, leads, as we will show in the following,
to the spontaneous formation of helical configurations.

Our model chromosomes are of the bead-spring type, with consecutive beads
attached to each other by a harmonic spring $V_b = (A/2)({\bf d_i} - \s  {\bf u_i})^2$ where
${\bf d_i} = \br_{i+1}-\br_i$, ${\bf r_i}$ is the position of $i$-th bead, $\s$ the equilibrium
bond-length and ${\bf u_i} = {\bf d_i}/|{\bf d_i}|$ is the local tangent vector to the chain.
Non-bonded beads repel each-other through the Weeks-Chandler-Andersen (WCA) potential \cite{weeks1971}
$V_{WCA} = 4 \e \left[ (\s/r_{ij})^{12}-(\s/r_{ij})^6+{1}/{4}\right]$ if
the inter-monomer separation $r_{ij} < 2^{1/6}\s$ else $V_{WCA} =0$,
where $\e$ and $\s$ set the energy and length scale of the system respectively.
We use $A=100 \e$.
The interaction of all beads with the confining walls are modelled through
$V_{wall} = 2\pi\e [(2/5)(\s/r_{iw})^{10}-(\s/r_{iw})^4 + 3/5]$ if the distance of the $i$-th monomer
from a wall $r_{iw} < \s$ and $V_{wall} =0$ otherwise. We simulate this system employing a
velocity-Verlet molecular dynamics scheme in presence of a Langevin
thermostat fixing the temperature at $\kb T=1$ as implemented by the ESPResSo package \cite{Limbach2006}.

%\subsection{Results}
%

We first consider a polymer composed of a linear backbone chain of length $l_b=200\s$ to which side-loops
of length $l_s=40 \s$ are attached at every backbone monomer of the main chain.
This polymer is confined to a cylinder of length $L=50.75 \s$ and diameter $D=29.5 \s$,
yielding a monomer packing fraction of $\n = 23.8\%$.
In Fig.~\ref{helix-1} we show a typical equilibrium configuration of this polymer,
which evidently displays a marked helical ordering of the backbone chain. The degree of helical ordering
can be quantified by considering the tangent-tangent correlation function
$\la {\bf u}(s)\cdot {\bf u}(0) \ra$, where the positional coordinate is given
by $s=i \s$ with $i=0,1, \ldots,200$. The Fourier transform of this quantity
yields a structure function $S(q)$ with a peak at a dimensionless wavenumber
$ q_{max}= l_b/\lambda_{max}$, where $\lambda_{max}$ is the pitch of the
helix measured along the backbone chain. The height of the structure function at its
maximum is a relative measure of the degree of helicity, whilst the width of the peak
is indicative of the statistical dispersion of the structure.

\begin{figure}[t]
\begin{center}
\includegraphics[width=8.6cm]{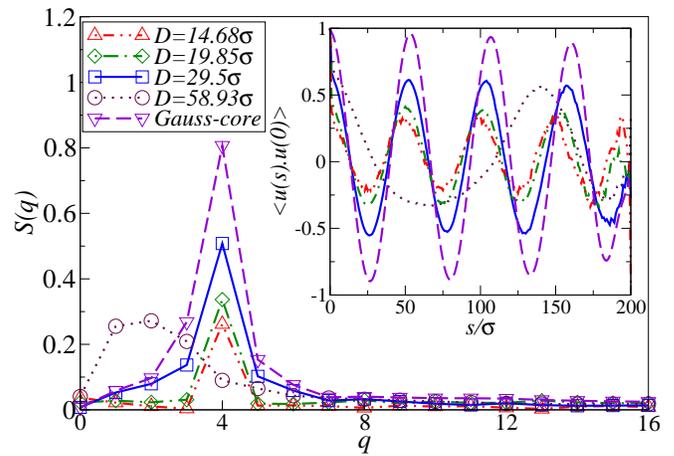} %{tan_cor.eps} %{r_tancorrln.eps}
%\vskip 1cm
%\includegraphics[width=8cm]{sk.eps}  %{r_sk-tancorln.eps}
\caption{(Color online)
The tangent-tangent correlation (inset) and its Fourier transform for the polymer shown in Fig.~\ref{helix-1}.
The correlation function is oscillatory, with the periodicity captured by the peak in the structure factor
at %$q_p=4$ for $D=58.93, 29.5, 19.85$, and $q_p=2$ for $D=14.68$.
$q_p=4$ for $D/\s=14.68, 19.85,  29.5$, and $q_p=2$ for $D/\s=58.93 $.
Also shown are the corresponding results for a main-chain polymer with an effective Gaussian-core mimicking the effect of inter-side-chain repulsion
(cf.  Fig.~\ref{wlc-helix}($b$)).}
\label{tan-helix-1}
\end{center}
\end{figure}

Fig.~\ref{tan-helix-1} shows the correlation function and the corresponding structure function for this polymer,
as we vary the diameter of the confining cylinder. This shows that the helical pitch is relatively robust against
changes in the diameter, although a slight decrease in the amplitude is apparent as we increase the diameter. Only for the largest diameter, when both the degree of confinement as well as the overall packing fraction are significantly decreased, do we see a preference for a more longitudinal packing of the main chain.

\begin{figure}[t]
\begin{center}
\includegraphics[width=8.6cm]{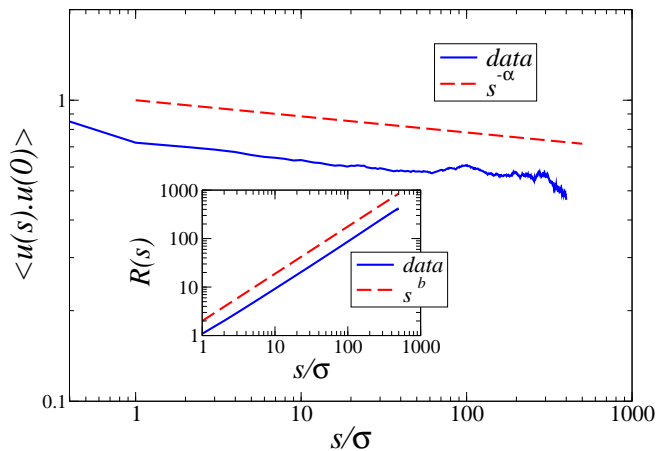}
\caption{(Color online) The tangent-tangent correlation
%$\la {\bf u}(0). {\bf u}(s)\ra$
 shows a weak power-law decay $s^{-\a}$ with
$\a=0.06$. Inset: power-law growth of distance between two segments
$R(s)=\sqrt{\la r(s)^2 \ra} \sim s^b$ with $b=0.97$.}
\label{stiff}
\end{center}
\end{figure}

We now argue that the helical arrangement is stabilized by two effects. The first is the effective stiffness induced in the backbone by
the presence of side loops. To quantify this intrinsic effect in the absence of confinement we study a \emph{free} polymer of length $l_b=500$,
with side-loops of the same length as before, $l_s=40 \s$, grafted at each backbone monomer.
Although for a very long backbone ($l_b \gg l_s$)
one expects the exponent $b$, which governs the mean-square separation between monomers at a distance $s$ along the backbone through
the scaling $\la r(s)^2 \ra \sim s^{2b}$, to reproduce the value $b = 3/5$ of a simple self-avoiding polymer, we find at the shorter length-scales
relevant to our confined polymer a much higher value of $b=0.97$ (Fig.\ref{stiff}, inset).
 At the same time, the tangent-tangent correlation $\la {\bf u}(s). {\bf u}(0) \ra$ (Fig.\ref{stiff}) shows an extremely weak power-law decay $s^{-\a}$
 with $\a=0.06$, which satisfies the generic scaling rule $\a=2-2b$. The small value of the exponent $\a \ll 1$ suggests that at these relatively short
 length scales the backbone is characterized by a significant effective stiffness, even close to that of a of a rigid rod for which $b=1$.
 The algebraic nature of the correlation decay, however, precludes a simple interpretation in terms of an intrinsic length, in contrast to the
 persistence length of a chain with intrinsic resistance to bending. Nevertheless, it is intuitively clear that there is a free-energy cost associated
 with local deformations of the backbone caused by the changes induced in the side-loop packing. For comparison,  we note that
 the end-to-end distance measured for our backbone with $l_b=500$, which we determined to be $R \approx 413\, \s$, would be reproduced by a
 worm-like chain with persistence length $P \approx 391\, \s$~\cite{Hermans1952} comparable to the contour length.

\begin{figure}[t]
\begin{center}
\includegraphics[width=4cm]{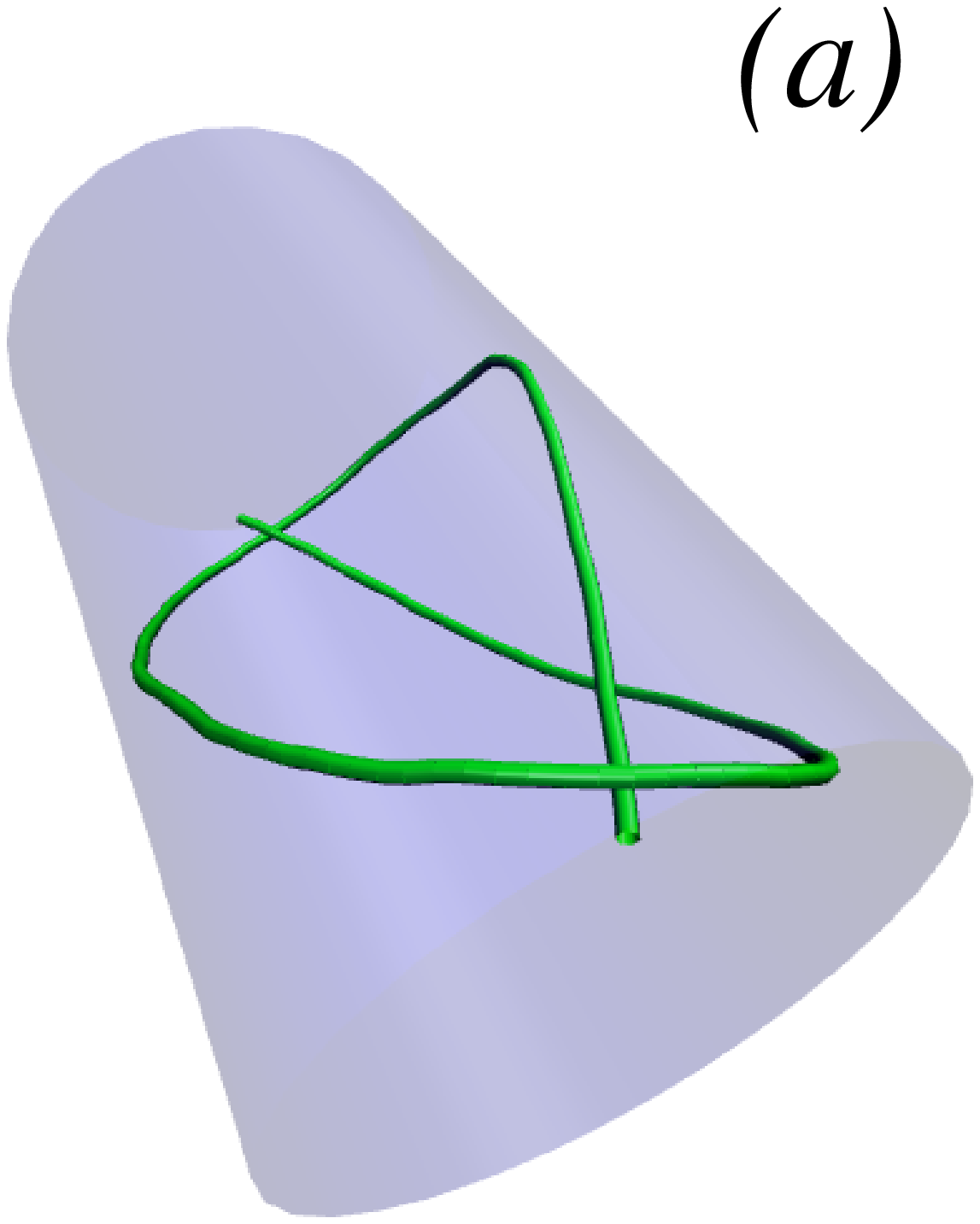} %{2wlc_cyl.eps} %{two_mol_in_cyl.eps} %{wlc-gc-compare.eps} %{wlc-helix.eps}
\includegraphics[width=4cm]{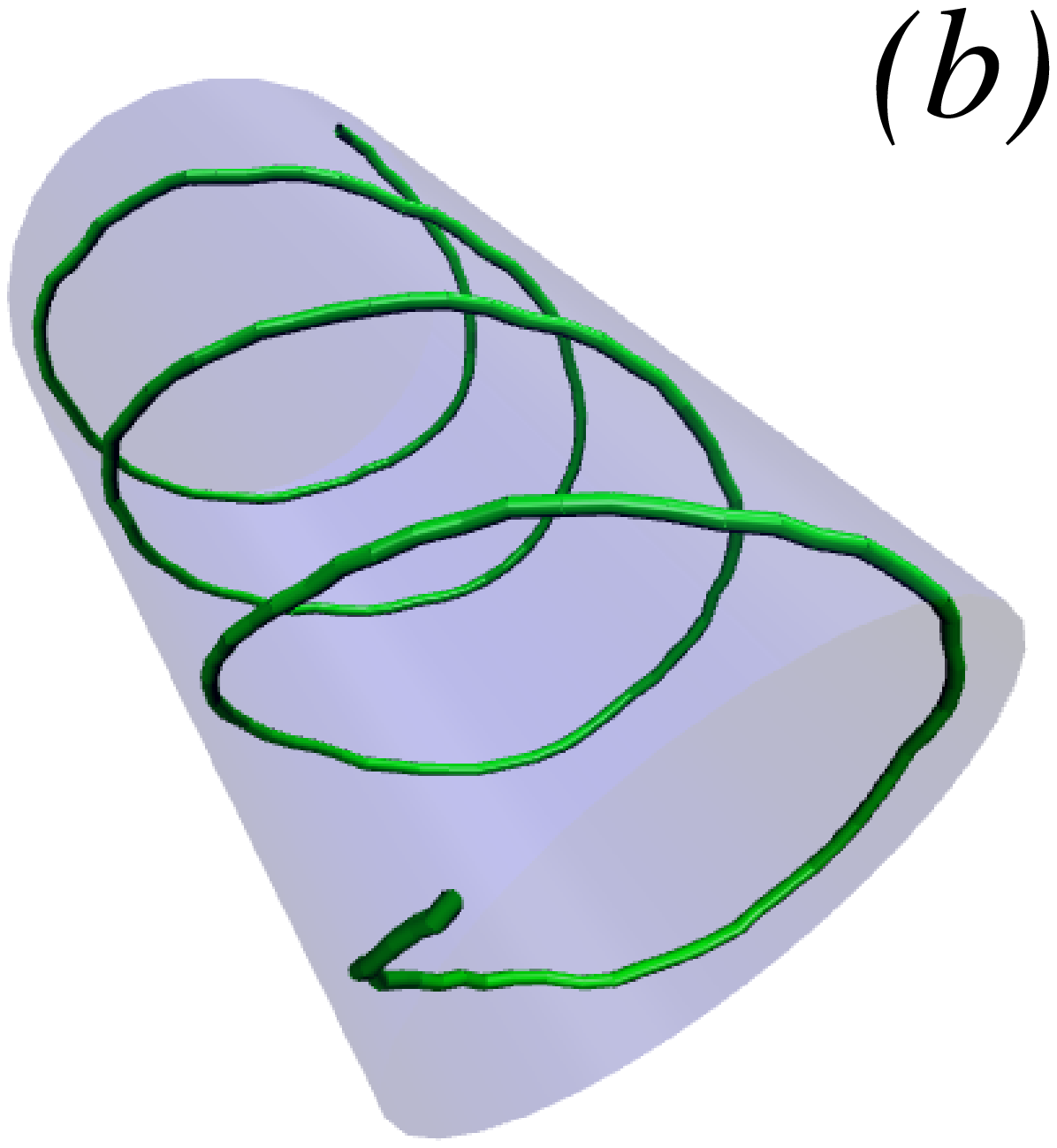} %{poly_gauss_molecule.eps}%{gcwlc_cyl.eps}
\caption{(Color online) $(a)$ Self-avoiding WLC and $(b)$ self-avoiding Gaussian-core polymer within a cylinder.
The chain has contour length $l_b=200 \s$, and for the WLC polymer the persistence length $P = 2 l_b$.
The confining cylinder is
of length $L=50.75 \s$ and diameter $D=29.5 \s$ for both $(a)$ and $(b)$.}
\label{wlc-helix}
\end{center}
\end{figure}

However, backbone stiffness alone is not sufficient to explain the emergent helicity.
Although earlier work has shown that a persistent chain confined to the \emph{surface}
of a cylinder can adopt helical confirmation \cite{Kusner2006}, a persistent chain confined to the \emph{volume}
of a cylinder does not. We confirm this by simulating a worm-like-chain (WLC) of length $l_b=200\, \s$ with a persistence
length of $P = 2 l_b$, close to the effective value determined above for the free polymer.
A typical equilibrium structure of the confined WLC is shown in Fig.~\ref{wlc-helix}, which clearly displays
the tendency of such a chain to align itself with the long-axis of the cylinder, without much internal
structure developing, also supported by an analysis of the tangent-tangent correlations (see supplementary material).

This negative result for the persistent chain, points to the importance of the packing of the
side-loops in stabilizing the helical structure. One could naively expect that the interactions
between the side loops cause an additional effective soft repulsion between the backbone monomers
with a range determined by the radius of gyration of the side-loops. This repulsion has a dual role;
as we have shown above, it stiffens the backbone at short length scales, but at long length scales it acts to keep distant parts of the backbone apart,
effectively thickening the backbone to a soft `tube'. To assess the validity of
this latter idea we consider a chain whose monomers, apart from interacting through the short range WCA potential, also interact with an effective
Gaussian core interaction
\beq
\be V_{gc}(r) = a \exp[-(r/w)^2]
\eeq
intended to mimic, as simply as possible, the soft repulsion between the side-loops.
Bolhuis et al.\ showed that, in free space, the effective interaction between two linear polymers
of the same length can be approximated by the above expression with $a \sim 2$  and
$w=R_g$, the radius of gyration of the individual polymers~\cite{Bolhuis2001}.
While it is well known that the free energy cost of overlap between two polymers in free space
is independent of polymer size~\cite{DesCloizeaux1975, Daoud1975, Grosberg1982}, the {\em blob}
picture of polymers suggests that their overlap free energy will become a linear function
of polymer length~\cite{Jun2006} when the chains are strongly confined.
We therefore take the interaction strength  $a=40$ to mimic the densely-packed side-loops of size $l_s=40\s$.
From our simulations of side-loop coupled back-bone, we found the mean radius of gyration $R_g=2.95\,\s$ of the side-loops of
length $l_s=40\s$. However, due to the repulsive interactions with the backbone monomers the center of mass of the
loops is offset from the backbone. We therefore extracted the distribution of the center of mass of the side-loops with respect
to their attachment point on the backbone in the plane perpendicular to the local tangent direction along the chain
(see  Figure~\ref{cm_hist} in the Supplementary Material). This distribution has a sharp maximum at a distance of approximately
$4\s$ from the backbone chain. This result also shows that indeed the density distribution of the side-loop monomers is structured in a manner consistent with the `thickened tube' picture alluded to above. We therefore took the size parameter characterizing the range of the interaction between the side-loops
in the effective potential centered on the backbone monomers $V_{gc}$ to be $w \simeq R_g+4\s = 6.95\s$. This effective interaction between the monomers was then added to
the WCA potential, governing the non-bonded repulsive interactions, for a
linear chain of length $l_b=200 \s$ confined within a cylinder of diameter $D=29.5\s$
and length $L=50.75\s$. As Fig.\ref{wlc-helix}b shows, this effective potential reproduces the
helical equilibrium structures of the polymer remarkably well. The structure factor displays a maximum at the same helical
pitch value $l_b/4=50\s$ as the original simulations and reproduces the oscillations of the tangent-tangent
correlation function (Fig.~\ref{tan-helix-1}) with only a slight global phase shift. The amplitude of the
oscillations (and hence that of the structure factor) is somewhat larger for the effective potential,
but this is probably due to an overestimate of the interaction strength $a$.

\begin{figure}[t] %[htbp]
\begin{center}
\includegraphics[height=6cm]{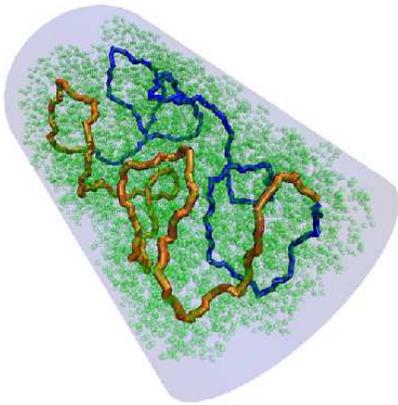} %{ringmol_with_cyl.eps} %{nu-helix.eps}
\caption{(Color) Equilibrium structure for a circular backbone chain of length $l_b=400\s$, with side-loops
of length $l_s=20 \s$  attached to each backbone monomer. The polymer is confined within a cylinder
of length $L=50.4 \s$ and diameter $D=33.5 \s$. The splitting of the backbone into two parallel
{\em linearized} branches is highlighted by using two color-codes, orange and blue, for the branches.
The side-loop monomers are shown as transparent green beads.}
\label{helix-2}
\end{center}
\end{figure}

Finally we address what happens if the backbone is a ring-like polymer,
appropriate to modeling, e.g., the circular chromosome of \emph{E. coli}. To that end we
simulated a polymer with a circular backbone with $l_b=400 \s$ and side-loops of
length $l_s = 20 \s$, trapped within a cylinder of
length $L=50.4 \s$ and diameter $D=33.5 \s$. The packing fraction of the monomers
is $\n = 18.9\%$, comparable to the one in the simulation of the linear backbone polymer.
Fig.~\ref{helix-2} shows that in this case the backbone loop is now organized into two parallel
helices running along the long axis of the cylinder. As is evident from the snapshot,
and corroborated by the analysis of the tangent-tangent correlations (Fig. \ref{tan-helix-2} in Supplementary Material), the degree of helicity is reduced as compared to the linear backbone case due to the smaller side-loop length, but nevertheless
remains significant.

In conclusion, we have shown that the interplay between the effective stiffness and intra-chain packing effects caused by side-loops in polymers leads to novel helical equilibrium configurations of confined polymers. These structures are strikingly similar to ones recently observed in bacterial nucleoids. To what extent the physical effects discussed here are sufficient to explain all the details of the large scale chromosome organisation in real bacteria is a question which clearly requires further research. At the very least, however, our results once again indicate that the ubiquitous aspecific interactions between the segments of long biopolymers like DNA can by themselves lead to significant spatial structuring, as has previously also been observed in the context of chromosome organisation in the nuclei of plants \cite{deNooijer2009} and humans \cite{Cook2009}. This argues for a more prominent place for polymer physics in the research into the structure and function of chromosomes. From a purely physical point of view our work points to novel possibilities for ``sculpting'' the configurations of confined polymers by judicial choices of polymer topologies.

\acknowledgments We gratefully acknowledge discussions with Nancy Kleckner and Mara Prentiss (Molecular and Cellular Biology, Harvard) in the initial stages of this project. This work is part of the research program of the
\textquotedblleft Stichting voor Fundamenteel Onderzoek der Materie
(FOM)\textquotedblright, which is financially supported by the
\textquotedblleft Nederlandse Organisatie voor Wetenschappelijk Onderzoek
(NWO)\textquotedblright. The work of DC was supported by FOM-programme Nr.~103 ``DNA in action: Physics of the genome''.
\bibliographystyle{apsrev4-1}

\clearpage
%\newpage

\section{Supplementary: Spontaneous helicity of a polymer with side-loops confined to a cylinder}

\begin{figure}[hbtp]
\begin{center}
\includegraphics[width=8cm]{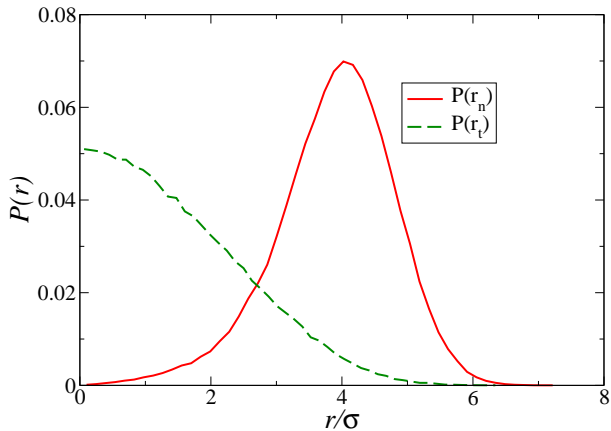} %{cm_hist.eps}
\caption{The distribution of center of mass distance of the side-loops around the main-chain.
$r_t$ denotes the component of the distance along the local tangent of the main chain and
$r_n$ denotes the component in a plane perpendicular to the local tangents. $P(r_n)$ has a pronounced 
maximum near $r_n=4\s$, whereas $P(r_t)$ has a Gaussian shape with a variance of $8\s^2$.}
\label{cm_hist}
\end{center}
\end{figure}

\begin{figure}[hbtp]
\begin{center}
\includegraphics[width=8cm]{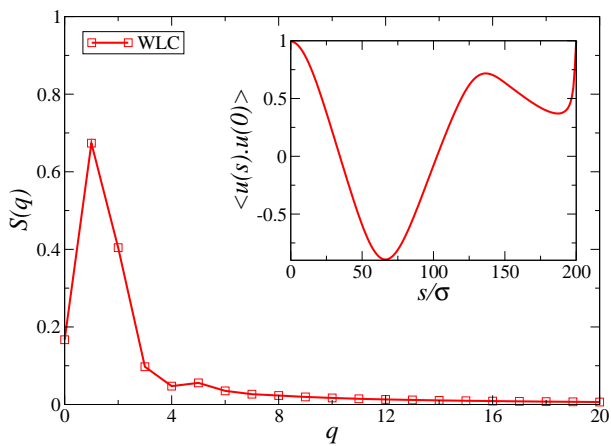} %{gc-wlc.eps}
\caption{The tangent-tangent correlation (inset) and its Fourier transform 
for a WLC polymer in presence of WCA repulsion between non-bonded monomers.}
\label{tan-wlc}
\end{center}
\end{figure}

\begin{figure}[hbtp]
\begin{center}
\includegraphics[width=8cm]{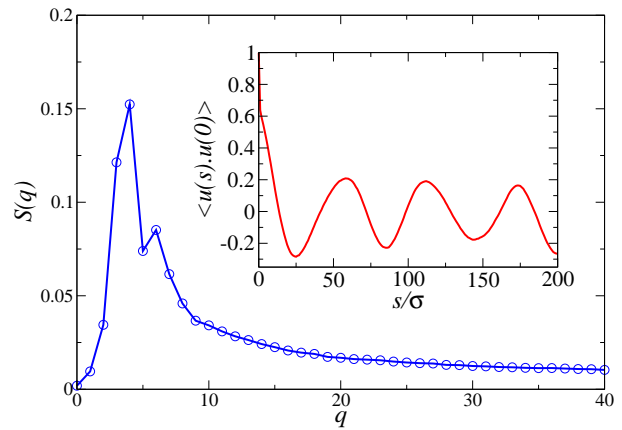}
\vskip 1.5cm
\caption{The tangent-tangent correlation and its Fourier transform for a circular main-chain polymer
attached with side-loops -- the main-chain has length $l_b=200\s$ and side-loops
of length $l_s=40 \s$ --
and confined within a cylinder of length $L=50.4 \s$ and diameter $D=33.5 \s$. The correlation (inset)
shows clear periodic oscillation, and the periodicity is captured by the peak in the Fourier transform
at $q_p=4$.}
\label{tan-helix-2}
\end{center}
\end{figure}

We calculate the probability distribution of the center of mass position of the side-loops 
around the main chain polymer (Fig.~\ref{cm_hist}). This is done using our direct simulation 
of a main-chain polymer attached with side-loops confined within a cylinder. 
The main-chain has length $l_b=200\s$ to each monomer of which is attached 
side-loops of length $l_s=40\s$. The polymer is confined within a cylinder of length $L=50.75 \s$ 
and diameter $D=29.5 \s$.

The center of mass position of the loop $\bf{r}$ is measured from the main-chain monomer to which 
the loop is attached. This position vector can be projected onto the direction 
of a local tangent on the main chain $\bf \hat t$, and a normal direction $\bf \hat n$
such that ${\bf r} = r_n {\bf \hat n}+ r_t {\bf \hat t}$, 
where $r_n = \bf r.\hat n$ and $r_t = \bf r. \hat t$. We calculate the distribution $P(r_t)$, $P(r_n)$
using $1000$ equilibrated configurations of the polymer (Fig.~\ref{cm_hist}). 

We find that the center of mass of the side chains are distributed in a Gaussian manner 
in the direction parallel to the local tangent $P(r_t)$. Howeever, in the perpendicular planes, 
the distribution $P(r_n)$ has a pronounced maximum  at a distance $r_n=4\s$. 
This off-center location of the center of mass of the side-loops, along with the radius of gyration 
$R_g=2.95\,\s$ leads us to use  $w\simeq R_g+4\s = 6.95\s$
for the effective inter-side-loop  interaction  $\be V_{gc}(r) = a \exp[-(r/w)^2]$.

In Fig.~\ref{tan-wlc} we plot the tangent-tangent correlation and its Fourier transform 
for a self-avoiding (non-bonded monomers repel by WCA potential) 
WLC polymer of chain-length $l_b=200\,\s$ and persistence length $P=2l_b$, confined within a cylinder of 
length $L=50.75\,\s$ and diameter $D=29.75\, \s$ (corresponding to Fig.4($a$) of main text).
Note the reduced magnitude and pitch of the helicity comapred to that of self-avoiding polymer
in presence of the Gaussian-core repulsion as shown in Fig.1 of main text.

In Fig.~\ref{tan-helix-2} we show the tangent-tangent correlation and its Fourier transform
for the case when the backbone is a ring-like polymer shown in Fig.5 of main text. 
The correlation function shows nice periodic 
oscillation which is captured by the maximum in the Fourier transform at $q_p=4$. This peak
corresponds to a helical pitch of $\l_{max}=l_b/4 = 50\,\s$.

\end{document}